\documentclass[fleqn,aps,pra,amsmath,amssymb,reprint,showpacs]{revtex4-1}
\usepackage{graphicx}
\usepackage{dcolumn}
\usepackage{bm}
\usepackage{epstopdf}
\usepackage{amsmath}
\usepackage{textcomp}
\usepackage{braket}

\begin{document}
\title{Measuring the linewidth of a stabilized diode laser}

 \author{Lal Muanzuala}
 \author{Harish Ravi}
 \author{Karthik Sylvan}
 \author{Vasant Natarajan}
 \affiliation{Department of Physics, Indian Institute of
 Science, Bangalore 560\,012, India}
 \email{vasant@physics.iisc.ernet.in}
 \homepage{www.physics.iisc.ernet.in/~vasant}

\begin{abstract}
We demonstrate a straight-forward technique to measure the linewidth of a grating-stabilized diode laser system---known as an external cavity diode laser (ECDL)---by beating the output of two independent ECDLs in a Michelson interferometer, and then taking the Fourier transform of the beat signal. The measured linewidth is the sum of the linewidths of the two laser systems. Assuming that the two are equal, we find that the linewidth of each ECDL measured over a time period of 2 \textmu s is about 0.3 MHz. This narrow linewidth shows the advantage of using such systems for high-resolution spectroscopy and other experiments in atomic physics.\\

\end{abstract}


\maketitle

\section{Introduction}
The advent of diode lasers in the last couple of decades has revolutionized laser spectroscopy in atoms, and made possible several experimental studies in fields such as precision measurements \cite{BDN03,DAN05,RCN11}, laser cooling and trapping of atoms \cite{RWN01}, atomic clocks \cite{BNW97}, quantum optics \cite{RWN03a,IKN08,CPB14}, and so on. This is because most experiments are done using the D lines of alkali atoms, which are in the near infrared (IR) and hence accessible with diode lasers. In addition, alkali atoms have high vapor pressure at room temperature so that vapor cells with sufficiently high atomic density can be used.

However, in order to be useful for high-resolution atomic spectroscopy (where transitions have linewidths of a few MHz \cite{DAN08}) the laser linewidth should be below 1~MHz. Since the linewidth of a commercial diode laser (of the kind that is used in CD players for example) is of the order of a few GHz, it is necessary to reduce this linewidth. The required reduction is typically achieved using optical feedback from a diffraction grating, in what is called the Littrow configuration. This also serves the purpose of making the frequency of the laser tunable by changing the angle of the grating. The grating is usually mounted on a piezoelectric transducer so that the angle can be changed electronically.

\begin{figure}
\centering{\resizebox{0.95\columnwidth}{!}{\includegraphics{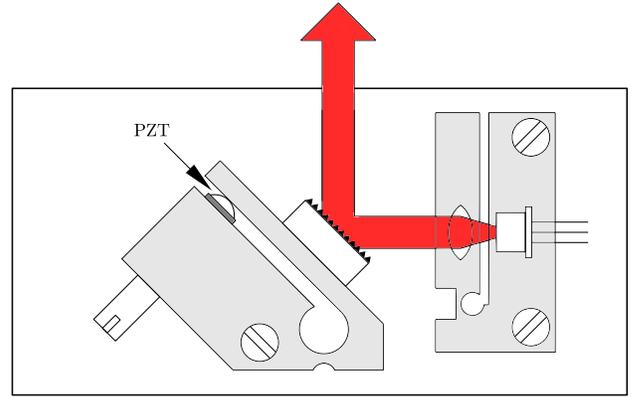}}}
\caption{(Color online) Diode laser stabilization in Littrow configuration. Optical feedback from a grating is used to reduce the linewidth of the laser. The grating is mounted on a piezoelectric transducer (PZT) to enable electronic tunning of the wavelength. This configuration is called an external cavity diode laser (ECDL).}
 \label{diodelaser}
\end{figure}

The configuration, shown schematically in Fig.\ \ref{diodelaser}, is arranged so that the $-1^{\rm th}$ diffraction order is fed back to the laser, while the specular reflection is the output. From the grating equation, we have
\begin{equation*}
2d \sin{\theta} = m \lambda
\end{equation*}
where $d$ is the spacing between the successive lines of the grating, and $ \theta $ is the angle of the $m^{\rm th}$ order diffraction. Since the specular reflection is the output beam, it is convenient to have $\theta$ close to $45^{\circ} $. This is achieved by choosing $d$ appropriately---e.g.\ the grating for accessing the D lines of K, Rb, and Cs (770--900 nm) has 1800 lines/mm. The power available after optical feedback is usually about $70\% $ of the open-loop power. Thus, although the linewidth reduction of the diode laser is by more than a factor of 1000, the loss in power is only $ 30\% $, showing that this is not wavelength selection (as for a grating used to disperse the light from a white-light source) but actual reduction in wavelength uncertainty. In effect, the grating along with the back facet of the diode forms a second lasing cavity---which is why this configuration is called an external cavity diode laser (ECDL)---with the longer cavity resulting in a smaller linewidth.

The linewidth of the laser can be measured in a Michelson interferometer, with the requirement that the phase difference between the two arms be larger than the phase coherence of the laser \cite{BOW99}. If the linewidth of the laser is about 1 MHz, the corresponding coherence length is 300~m. This means that the two arm-lengths of the interferometer have to differ by a kilometer or more. This is not easy to implement in the lab unless one uses a coiled optical fiber of that length. An alternate (and easier) way would be to use two identical laser systems and interfere/beat them in the interferometer. Since the phase of the two lasers are independent, the two arm-lengths can be nominally equal, with the understanding that the beat signal will represent the convolution of the two laser linewidths. If we assume that the two ECDLs have Lorentzian distributions with center frequencies $\omega_1$ and $\omega_2$, and linewidths (full-width-at-half-maximum, FWHM) $ \Gamma_1$ and $\Gamma_2$, respectively, then the normalized distribution in frequency-space is 
\begin{equation}
\label{eqbeat}
L_i(\omega) = \dfrac{1}{\pi} \dfrac{\Gamma_i/2}{(\omega-\omega_i)^2 + (\Gamma_i/2)^2}
\end{equation}
where $i$ is 1 or 2 for the two laser systems.


In this study, we present the results of such a linewidth measurement on two \textbf{home-built} ECDLs beat over a time period of 2 \textmu s. As expected, the linewidth of each laser is below 0.5 MHz. To see if there is an effect of locking the frequency of a laser, we have done three studies---(i) both lasers free running, (ii) one laser locked and the other free running, and (iii) both lasers locked. The results indicate that locking lasers has no effect on the linewidth, at least over this time scale.

\section{Experimental details}

The diode laser system consists of a Sharp laser diode (GH0781JA2C) operating with a free-running wavelength of 784 nm and power of 120 mW. The laser is stabilized using feed back from an angle-tuned grating with 1800 lines/mm, as shown in Fig.\ \ref{diodelaser}. The system is mounted on a thermo-electric cooler for temperature stabilization. Using a combination of operating temperature and operating current, the laser system is brought near the Rb D$_2$ line (5S$_{1/2}$ $ \rightarrow $ 5P$_{3/2}$ transition) at 780 nm. Part of the laser output is fed into a saturated absorption spectrometer (SAS) \cite{BOW99}, so that the laser can be locked to a hyperfine transition, if necessary. The locking is achieved by modulating the laser current at 20 kHz, and demodulating the SAS signal using a lock-in amplifier.

\begin{figure}
\centering{\resizebox{0.95\columnwidth}{!}{\includegraphics{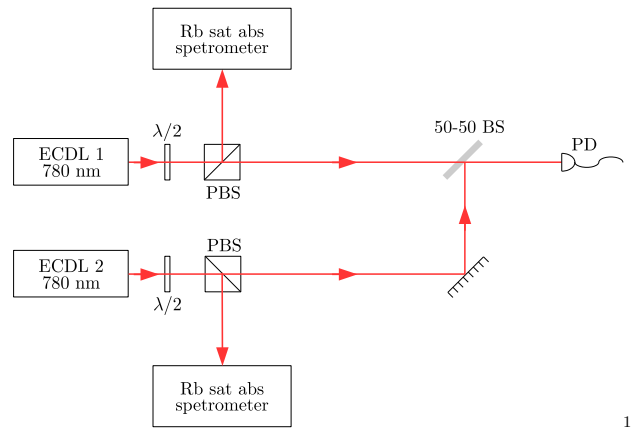}1}}
\caption{(Color online) Experimental schematic for the beat signal measurement. Figure key: $ \lambda/2 $ -- halfwave retardation plate, PBS -- polarizing beam splitter cube, BS -- beam splitter, PD -- photodiode.}
 \label{beatschematic}
\end{figure}

The experimental setup for obtaining the beat signal is shown in Fig.\ \ref{beatschematic}. The two lasers have a fixed frequency difference of about 16 MHz. This is so that the beat signal is centered around a non-negative value, and the variation around this value can be measured unambiguously. By contrast, if the frequency difference were zero, the lineshape would be a half-Lorentzian function, because only positive frequencies would appear in the spectrum. The output of the two ECDLs is mixed on a 50-50 non-polarizing beam splitter, as shown in the figure. The beat signal is measured on a fast photodiode, with response time sufficiently fast in order to measure the 16 MHz signal. The signal is measured at a sampling rate of 1 GHz for a total time of 2 \textmu s, corresponding to 2000 points. A fast Fourier transform (FFT) of the signal gives the frequency spectrum, with sufficient zero padding to make the spectrum smooth.

\section{Results and discussion}

\begin{figure}
\centering{\resizebox{0.95\columnwidth}{!}{\includegraphics{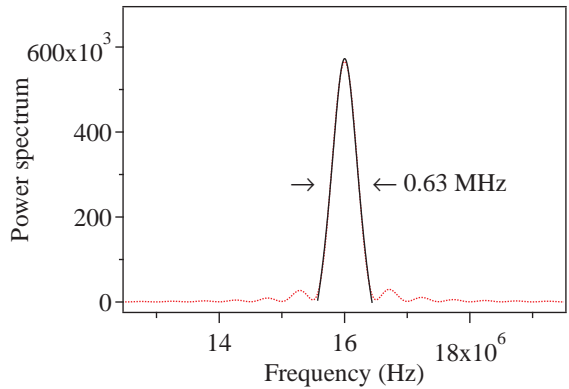}}}
\caption{(Color online) Calculated power spectrum for a function given by Eq.\ \eqref{eqfunc} with center frequency 16 MHz, linewidth 0.6 MHz, and lasting for a time of 2 \textmu s (shown with a dotted line). The solid line is a Lorentzian fit to the central peak, which matches the spectrum almost perfectly and yields a linewidth of 0.63 MHz.}
 \label{calcfft}
\end{figure}

Before turning to the experimental results, we see what is the expected lineshape for the beat signal. A Lorentzian centered at non-zero frequency  can be simulated using a function of the form 
\begin{equation}
\label{eqfunc}
f(t) = e^{-2\pi \gamma t/2} \cos (2 \pi f_\circ t)
\end{equation}
where $\gamma$ is the linewidth, and $f_\circ $ is the center frequency. We take typical values of $\gamma = 0.6$ MHz and $f_\circ = 16 $ MHz. Using experimental values of 2000 samples at a sampling rate of 1 GHz and total time of 2 \textmu s, the FFT of this function (magnitude squared with zero padding of 100000 points) is shown in Fig.\ \ref{calcfft}. The lineshape is essentially the convolution of a Lorentzian with a sync function (because of the finite time duration of the function) as evidenced by the zeros of the spectrum. The lineshape near the peak is mainly Lorentzian as seen from the near-perfect overlap with the Lorentzian fit, and the linewidth obtained from the fit is 0.63 MHz---close to the chosen value of 0.6 MHz. Therefore, in the following, the experimentally measured spectrum is fit to a Lorentzian, and the linewidth determined from the fit.

\begin{figure}
\centering{\resizebox{0.95\columnwidth}{!}{\includegraphics{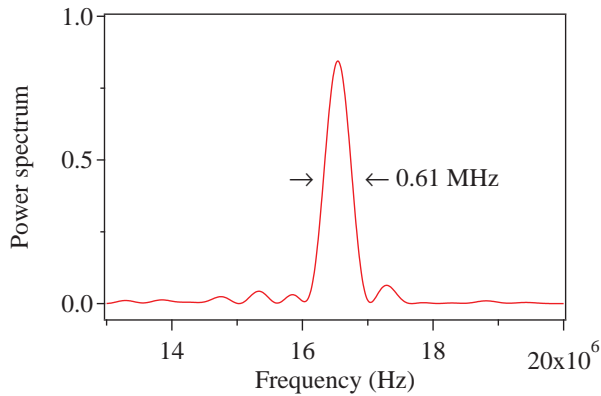}}}
\caption{(Color online) Experimental power spectrum obtained by beating two ECDLs. A Lorentzian fit (not shown) yields a linewidth of 0.64 MHz.}
 \label{exptfft}
\end{figure}

A typical experimental FFT spectrum obtained with two free running lasers is shown in Fig.\ \ref{exptfft}. The data are taken at a sampling rate of 1 GHz and for a total time of 2 \textmu s---exactly the conditions used for the theoretical results presented in Fig.\ \ref{calcfft}. It has a similar lineshape with zero points due to the finite signal duration. A Lorentzian fit to the central peak yields a linewidth of 0.61 MHz. If we assume that the two lasers are identical, then Eq.\ \eqref{eqbeat} shows that the linewidth of each laser is 0.3 MHz. Since this is the linewidth obtained after 2 \textmu s, it can be regarded as an average linewidth over this period. The instantaneous linewidth is expected to be lower.

In order to study the effect of locking the laser to a hyperfine transition, we have repeated the above experiment with one of the lasers locked, and with both the lasers locked. The results for all three sets are listed in Table \ref{table1}. Each measurement was repeated three times and what is listed is the average value.  There is not much difference in the values, which tells us that the linewidth does not change because of locking, \textit{at least over the 2 \textmu s timescale of the measurement}. We expect that the effect of locking will be to prevent long-term drift of the laser frequency, which can be important in experiments like laser cooling and quantum optics.

Noting that there is not much change in the average value over the three sets, and in order to have sufficient points to get a meaningful standard deviation, all nine measurements were combined into one set and the standard deviation calculated for the entire set---this value is 0.059 MHz, and can be regarded as an error bar on the linewidth measurement.

\begin{table}
\caption{Measured linewidths of the beat signal under different conditions of locking of the two ECDLs. Listed is the average value from 3 measurements.}
\label{table1}
\begin{tabular}{lc}
\hline \hline \\[-1ex]
Condition of ECDLs & Average linewidth (MHz) \\ [3mm] \hline \\[-1ex]
Both free running & $0.57$ \\
One locked and one free running & $0.60$ \\
Both locked & $0.58$ \\ [3mm]
\hline \hline
\end{tabular}
\end{table}

\section{Conclusions}

In conclusion, we have presented a technique where the linewidth of a grating-stabilized diode laser can be measured using a Michelson interferometer. Instead of the usual technique of having a km long fiber in one of the arms of the interferometer to create the required phase delay, we use the simpler technique of having two independent diode lasers with nominally equal path lengths in the two arms. If we assume that the linewidths of the two laser systems are equal, we find that the linewidth averaged over 2 \textmu s is about 0.3 MHz. This shows the advantage of using such stabilized diode laser systems (ECDLs) for high-resolution spectroscopy and other experiments in atomic physics.

\begin{acknowledgments}
This work was supported by the Department of Science and Technology, India.
\end{acknowledgments}


\end{document}